\documentclass[aps,pra,twocolumn,showpacs,groupedaddress,letterpaper,nofootinbib]{revtex4-1}

\usepackage{graphicx}
\usepackage{subfigure}
\usepackage{color}
\usepackage{physymb}
\usepackage{enumitem}
\usepackage{mdframed}
\usepackage{framed}
\usepackage{epstopdf}
\usepackage{natbib}
\usepackage{hyperref}
\usepackage{url}
\usepackage{multirow}
\usepackage{cleveref}
\usepackage{tabularx}
\usepackage{array}
\usepackage{float}

\newcolumntype{x}[1]{%
>{\centering\hspace{0pt}}p{#1}}%

\begin{document}

\title{Teaching Quantum Interpretations: Revisiting the goals and practices of introductory quantum physics courses}


\author{Charles Baily}
\affiliation{School of Physics and Astronomy, University of St Andrews, St Andrews, Fife KY16 9SS Scotland, UK}

\author{Noah D. Finkelstein}
\affiliation{Department of Physics, University of Colorado, 390 UCB, Boulder, CO 80309 USA}

\begin{abstract}Most introductory quantum physics instructors would agree that transitioning students from classical to quantum thinking is an important learning goal, but may disagree on whether or how this can be accomplished.  Although (and perhaps because) physicists have long debated the physical interpretation of quantum theory, many instructors choose to avoid emphasizing interpretive themes; or they discuss the views of \textit{scientists} in their classrooms, but do not adequately attend to \textit{student} interpretations.  In this synthesis and extension of prior work, we demonstrate: (1) instructors vary in their approaches to teaching interpretive themes; (2) different instructional approaches have differential impacts on student thinking; and (3) when student interpretations go unattended, they often develop their own (sometimes scientifically undesirable) views.  We introduce here a new modern physics curriculum that explicitly attends to student interpretations, and provide evidence-based arguments that doing so helps them to develop more consistent interpretations of quantum phenomena, more sophisticated views of uncertainty, and greater interest in quantum physics.  
\end{abstract}

\maketitle

\section{\label{sec:intro}Introduction}

\let\thefootnote\relax\footnote{\textit{Lorem ipsum dolor sit amet, consectetur adipiscing elit. Ut lectus nisl, vestibulum et rutrum vel, hendrerit nec dolor. Etiam dignissim, augue ut faucibus vehicula, neque lacus consectetur neque, ac ullamcorper leo tellus ac eros. Maecenas. }}

\begin{quote}{``Why do some textbooks not mention \textit{Complementarity}?  Because it will not help in quantum mechanical calculations or in setting up experiments.  Bohr's considerations are extremely relevant, however, to the scientist who occasionally likes to reflect on the meaning of what she or he is doing.''

{-} Abraham Pais \cite{pais1991times}} \end{quote}

There have been numerous studies of student reasoning and learning difficulties in the context of quantum physics \cite{fischler1992modern,johnston1998quantum,mannila2002entities,thacker2003models,wuttiprom2009survey,zhu2012qm1,didis2014mental}, as well as related efforts to transform instructional practices so as to improve learning outcomes \cite{mckagan2006reform,mckagan2008phet,deslauriers2011quantum,zhu2012qm2}.  However, relatively little attention has been paid to the intersection of mathematics, conceptual framing and classroom practices, and how these impact students' understanding of quantum phenomena \cite{mueller2002teaching,morgan2006thesis,baily2011thesis}.

In education research, the term \textit{hidden curriculum} generally refers to aspects of science and learning that students develop attitudes and opinions about, but are primarily only implicitly addressed by instructors \cite{redish1998expect}.  Students may hold a variety of beliefs regarding the relevance of course content to real-world problems, the coherence of scientific knowledge, or even the purpose of science itself, depending (in part) on the choices and actions of their instructors.  Research has demonstrated that student attitudes tend to remain static or become less expert-like when instructors do not explicitly attend to them \cite{redish1998expect,adams2006class}.

The physical interpretation of quantum theory has always been a controversial topic within the physics community, from the Bohr{-}Einstein debates \cite{einstein1935epr,bohr1935epr} to more recent disagreements on whether the quantum state is \textit{epistemic} or \textit{ontic} \cite{harrigan2010epistemic,pusey2012reality}.  Although physicists have historically, as part of the discipline, argued about the nature of science, and the relationship between mathematical representations and the physical world, there is a fairly common tendency for instructors to de-emphasize the interpretive aspects of quantum mechanics in favor of developing proficiency with mathematical tools.  At the same time, other instructors may highlight the views of \textit{scientists} in their classrooms, but do not adequately attend to \textit{student} interpretations.

In other words, interpretation is typically a hidden aspect of quantum physics instruction, in the following sense: (a) it is often treated superficially, in ways that are not meaningful for students beyond the specific contexts in which the discussions take place; (b) students will develop their own ideas about quantum phenomena, particularly when instructors fail to attend to them; and (c) student interpretations tend to be more novice-like (intuitively classical) in contexts where instruction is less explicit \cite{baily2010teaching,baily2010hidden}.

This paper synthesizes and extends prior work \cite{baily2009development,baily2010teaching,baily2010refined,baily2010hidden,baily2011interpret} to provide evidence-based arguments for an instructional approach that emphasizes the physical interpretation of quantum mechanics.  To be clear, we are not advocating for more discussions of Schr{\"o}dinger's Cat in the classroom, but rather a greater emphasis on (for example) providing students with the conceptual tools and language to identify and articulate their own intuitions and beliefs about the classical world; and presenting them with experimental evidence that unambiguously challenges those assumptions.  We are also arguing for a re-evaluation of the usual learning goals for introductory quantum physics courses, so that mathematical tools are developed alongside conceptual understanding, rather than emphasizing calculation with the hope that students eventually come to understand what the quantum state might actually represent.

We present below an analysis of student data demonstrating the differential impact on student thinking of three different approaches to teaching interpretive themes in quantum mechanics.  One of the key findings is that students can be influenced by explicit instruction, but they frequently default to an intuitively classical perspective in a context where instruction was less explicit.  These results have motivated the development of a research-based modern physics curriculum that attends to student interpretations throughout the course.  We provide a summary overview of this curriculum, and present comparative studies demonstrating that our students developed more consistent interpretations of quantum phenomena, more sophisticated views of uncertainty, and greater interest in quantum physics.  We then revisit some of the reasons instructors choose to de-emphasize quantum interpretations, and discuss the broader implications of these choices for our students.

\section{\label{sec:background}Background and Courses Studied}

The University of Colorado Boulder (CU) offers two versions of its calculus-based modern physics course each semester: one section for engineering students, and the other for physics majors.  Both are delivered in large-lecture format (N$\sim$50$-$150), and typically cover the same general topics, spending roughly a quarter of the 15-week semester on special relativity, and the rest on introductory quantum mechanics and applications.  We have presented data from both types of courses in prior work \cite{baily2010teaching,baily2010hidden}, but every course to be discussed in this article is of the engineering kind, so that meaningful comparisons can be made between similar student populations.

In 2005, a team from the physics education research (PER) group at CU introduced a transformed curriculum for the engineering course that incorporated interactive engagement techniques (clicker questions, peer instruction and computer simulations), and emphasized reasoning development, model building, and connections to real-world problems \cite{mckagan2006reform}.  This new curriculum did not include relativity because the engineering faculty at CU felt that mechanical and electrical engineering students would benefit from learning more about modern devices and the quantum origin of material structure.  These course transformations, first implemented during the 2005/6 academic year, were continued in the following year by another PER group member (author:NF).  Subsequent instructors used many of these course materials and instructional strategies, but returned to including relativity in the curriculum.

\subsection{\label{instructors}Characterization of Instructional Approaches}

Our initial studies collected data from modern physics courses at CU during the years 2008-10.  With respect to interpretation, the instructional approach for each of these courses can be characterized as being either \textit{Realist}/\textit{Statistical}, \textit{Matter-Wave} or \textit{Copenhagen}/\textit{Agnostic}.  These characterizations are based on classroom observations, an analysis of course materials, and interviews with the instructors; they are not necessarily reflective of each instructor's personal interpretation of quantum physics, but rather whether and how they attended to interpretive themes in their teaching.  In this section, we focus on three individual instructors (A, B \& C), each of whom is representative of one of the three categories named above, as described in detail below.

These categories certainly do not encompass all the ways instructors might teach quantum interpretations, but they can be reasonably applied to every modern physics offering at CU during this time period, and we anticipate that most readers who have taught introductory quantum mechanics will recognize some similarity between their own approaches and those described below.  We are aware of other perspectives on teaching quantum physics that do not fit within these categories \cite{hobson2005field,rosenblum2011enigma,cheong2012suspense,norsen2013pilotwave}, but there are no published studies of their respective impacts on student learning; and still more interpretations of quantum theory exist \cite{cramer1986transaction,schlosshauer2005decohere,wallace2012everett,okon2014consistent,fuchs2014qb}, but we do not know of any literature describing their use in the classroom.

\begin{figure}[t]
    \includegraphics[height=0.8in, width=3.5in]{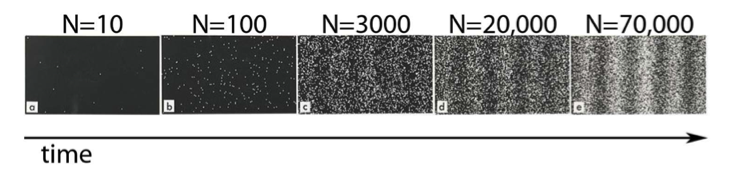}
\caption{Buildup of an electron interference pattern.  Single electrons are initially detected at seemingly random places, yet an interference pattern is still observed after detecting many electrons \cite{tonomura1989electron}.}\label{fig:electronbuildup}
\end{figure}

These different approaches to teaching interpretation can be best illustrated by how each instructor discussed the double-slit experiment with single electrons, though we have also taken into account instances in other contexts, and the frequency of such discussions throughout the semester \cite{baily2010teaching}. When this experiment is performed with a low-intensity beam, each electron will register individually at the detector, yet an interference pattern will still be seen to develop over time \cite{tonomura1989electron,frabboni2012buildup}. [See Fig.\ \ref{fig:electronbuildup}.]  Interference is a property associated with waves, whereas localized detections indicate a particle-like nature.  Different instructors will teach different interpretations of this result to their students, depending on their personal and pedagogical preferences. \newline

\noindent \textbf{Realist/Statistical (R/S):} Instructor A told students that each electron must pass through one slit or the other, but that it is impossible to determine which one without destroying the interference pattern.  Beyond this particular context, he also explained that atomic electrons always exist as localized particles, and that quantized energy levels represent the average behavior of electrons (because they are found to have a continuous range of energies when the measurement timescale is short compared to the orbital period, as enforced by the uncertainty principle).  During class, Instructor A referred to this as his own interpretation of quantum mechanics, one that other physicists might disagree with, and there was no discussion of alternatives to the perspective he was promoting.

To clarify, the label Realist/Statistical is being used here to denote a perspective wherein quanta exist as localized particles at all times, and the quantum state only encodes probabilities for the outcomes of measurements performed on an ensemble of identically prepared systems \cite{ballentine1970statistical}.  This is somewhat different from the purely statistical interpretation described by M{\"u}ller and Weisner \cite{mueller2002teaching}, who emphasized in their course that ``...classically well-defined dynamic properties such as position, momentum or energy cannot always be attributed to quantum objects.''

This local and realist perspective aligns with the na\"{i}ve interpretations that many introductory students construct when first trying to make sense of quantum phenomena.  Although it is less favored than other interpretations with regard to instruction, it does have its advocates.  For example, L. E. Ballentine uses the double-slit experiment in the introductory chapter of his graduate textbook to motivate an \textit{ensemble} interpretation of quantum mechanics: \begin{quote}``When first discovered, particle diffraction was a source of great puzzlement.  Are `particles' really `waves'?  In the early experiments, the diffraction patterns were detected holistically by means of a photographic plate, which could not detect individual particles.  As a result, the notion grew that particle and wave properties were mutually incompatible, or \textit{complementary}, in the sense that different measurement apparatuses would be required to observe them.  That idea, however, was only an unfortunate generalization from a technological limitation.  Today it is possible to detect the arrival of individual electrons, and to see the diffraction pattern emerge as a statistical pattern made up of many small spots.  Evidently, quantum particles are indeed particles, but particles whose behavior is very different from what classical physics would have led us to expect.'' \cite{ballentine1998modern}\end{quote}

Ballentine \textit{assumes} that localized detections imply the electrons were localized throughout the experiment, always passing through one slit or the other, but not both.  He explains diffraction patterns in terms of a quantized transfer of momentum between a localized particle and a periodic object. \newline

\noindent \textbf{Matter-Wave (MW):} From a \textit{Matter-Wave} perspective, the wave function is (for all intents and purposes) physically real: each electron \textit{is} a delocalized wave as it propagates through both slits and interferes with itself; it then randomly deposits its energy at a single point in space when it interacts with the detector.  The \textit{collapse of the wave function} is viewed as a process not described by the Schr\"{o}dinger equation, in which the electron physically transitions from a delocalized state (wave) to one that is localized in space (particle) \cite{bassi2013collapse}.

This is how Instructor B described this experiment during lecture, though he did not frame this discussion in terms of scientific modeling or interpretation, but rather presented students with (what he considered to be) sufficient experimental evidence in support of this view.  As he explained in a post-instruction interview:

\begin{quote}``This image that [students] have of this [probability] cloud where the electron is localized, it doesn't work in the double-slit experiment. You wouldn't get diffraction.  If you don't take into account both slits and the electron as a delocalized particle, then you will not come up with the right observation, and I think that's what counts. The theory should describe the observation appropriately.'' \end{quote}

Instructor B devoted classtime to interpretive themes at the beginning and very end of the quantum physics section of his course, but much less so in between (e.g., when teaching the Schr{\"o}dinger atomic model), with the presumption that students would generalize these ideas to other contexts on their own.  Of the various courses discussed in this paper, the quantum physics portion of Instructor B's course is the most similar to the original transformed curriculum developed in 2005. \newline

\noindent \textbf{Copenhagen/Agnostic (C/A):} The standard \textit{Copenhagen} interpretation \cite{faye2008copenhagen} would say this experiment reveals two sides of a more abstract whole; an electron is neither particle nor wave.  The dual use of (classically) distinct ontologies is just a way of understanding the behavior of electrons in terms of more familiar macroscopic concepts.  A wave function is used to describe electrons as they propagate through space, and the \textit{collapse postulate} is invoked to explain localized detections, but any switch between `particle' and `wave' occurs only in terms of how the electron is being represented.  The wave function is nothing more than a mathematical construct used to make predictions about measurement outcomes, without reference to any underlying reality.

Instructor C stated that a quantum mechanical wave of probability passed through both slits, but that asking which path an individual electron took without placing a detector at one of the slits is an ill-posed question at best.  The instructional emphasis for this topic was on calculating features of the interference pattern (determining the locations of maxima and minima), rather than physically interpreting the results.  This mostly pragmatic approach to instruction is also exemplified by a quote from a \textit{different} instructor (in a class for physics majors), who was asked during lecture whether particles have a definite but unknown position, or have no definite position until measured:

\begin{quote}``Newton's Laws presume that particles have a well-defined position and momentum at all times.  Einstein said we can't know the position.  Bohr said, philosophically, it has no position.  Most physicists today say: \textit{We don't go there. I don't care as long as I can calculate what I need.}''\end{quote}

The terms Copenhagen and Agnostic are being used jointly here to denote an instructional approach that is consistent with the Copenhagen interpretation, but de-emphasizes the interpretative aspects of quantum theory in favor of its predictive power (``Shut up and calculate!'' \cite{mermin1989pillow}); this should not to be confused with giving students a formal introduction to Bohr's stance on \textit{complementarity} and \textit{counterfactual definiteness}.

The purpose of this paper is not to debate the relative merits of these interpretations, but rather to explore the pedagogical implications of their use in the classroom.  Some key points to keep in mind are that the Realist/Statistical approach treats quantum uncertainty as being due to classical ignorance, and is aligned with students' intuitions from everyday experience and prior instruction.  From a Matter-Wave perspective, quantum uncertainty is a fundamental consequence of a stochastic reduction of the state upon interaction with a measurement device.  A Copenhagen/Agnostic instructor may regard quantum uncertainty as being fundamental, but generally considers such issues to be metaphysical in nature.

\subsection{\label{data}Initial Data Collection and Results}

At the beginning and end of most of the modern physics courses offered at CU during this time period, students were asked to fill out an online survey designed to probe their interpretations of quantum phenomena.  The survey consisted of a series of statements, to which students responded using a 5-point Likert scale (from \textit{strong agreement} to \textit{strong disagreement}); an additional textbox accompanied each statement, asking them to provide the reasoning behind their responses.  In this paper, the \textit{agree} and \textit{strongly agree} responses have been collapsed into a single category (agreement), and similarly for \textit{disagree} and \textit{strongly disagree}.

Students were typically offered nominal extra credit for completing the survey, or it was assigned in a homework set with the caveat that full credit would be given for providing thoughtful answers, regardless of the actual content of their responses.  The beginning of the survey emphasized that we were asking students to express their own beliefs, and that their specific answers would not affect any evaluation of them as students.  A few of the modern physics instructors were reluctant to provide academic credit for completing the survey; response rates from those courses were too low to be of use.

Some of the survey statements have evolved over time, primarily in the early stages of our research.  Modifications were generally motivated by a fair number of students providing reasoning that indicated they were not interpreting the statements as intended.  We conducted validation interviews with 19 students in 2009 \cite{baily2010refined}, after which the phrasing has remained essentially unchanged.  The student data presented in this paper were all collected from modern physics courses for engineers after the validation interviews took place.

An additional essay question at the end of the post-instruction survey presented statements made by three \textit{fictional} students regarding their interpretation of how the double-slit experiment with single electrons is depicted in the PhET Quantum Wave Interference simulation \cite{phetqwi} (as shown in Fig. \ref{fig:phetqwi}):

\begin{figure}[b]
    \includegraphics[height=1.5in, width=3.5in]{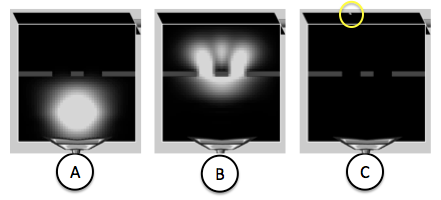}
\caption{A sequence of screen shots from the Quantum Wave Interference PhET simulation \cite{phetqwi}: (A) a bright spot emerges from an electron gun; (B) passes through both slits; and (C) a single electron is detected on the far screen (highlighted in this figure by the circle). After many electrons, a fringe pattern develops (not shown).}\label{fig:phetqwi}
\end{figure}

\begin{quote}
\textbf{Student 1:} The probability density is so large because we don't know the true position of the electron. Since only a single dot at a time appears on the detecting screen, the electron must have been a tiny particle, traveling somewhere inside that blob, so that the electron went through one slit or the other on its way to the point where it was detected.

\textbf{Student 2:} The blob represents the electron itself, since an electron is described by a wave packet that will spread out over time. The electron acts as a wave and will go through both slits and interfere with itself. That's why a distinct interference pattern will show up on the screen after shooting many electrons.

\textbf{Student 3:} Quantum mechanics is only about predicting the outcomes of measurements, so we really can't know anything about what the electron is doing between being emitted from the gun and being detected on the screen.
\end{quote}

\begin{figure}[b]
   \includegraphics[height=5.1cm, width=8.5cm]{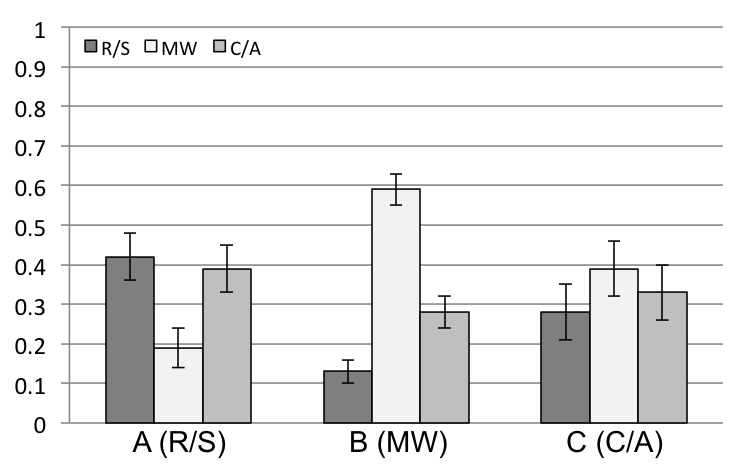}
\caption{Post-instruction student responses to the double-slit essay question for courses A, B \& C, where the labels R/S, MW and C/A refer to the instructional approach of each course, but also to each of the three statements in the essay question for which students expressed a preference.  N = 64 (A), 133 (B) \& 46 (C); error bars represent the standard error on the proportion.}\label{fig:dsessay}
\end{figure}

Respondents were asked to state which students (if any) they agreed with, and to explain their reasoning.  Generally speaking, aggregate responses for individual courses were similar to other courses that fell within the same category (R/S, MW or C/A).  Focusing on just the three courses described above, Instructor A's students were as likely to express a preference for the R/S statement (Student 1) as they were to prefer the C/A stance (Student 3); they were also the least likely group to prefer the MW description (Student 2).  Over half of Instructor B's students aligned themselves with the MW perspective on this experiment, whereas Instructor C's students were (within statistical error) evenly split among the three. [Fig. \ref{fig:dsessay}.]

\begin{figure}[t]
   \includegraphics[height=5.1cm, width=8.5cm]{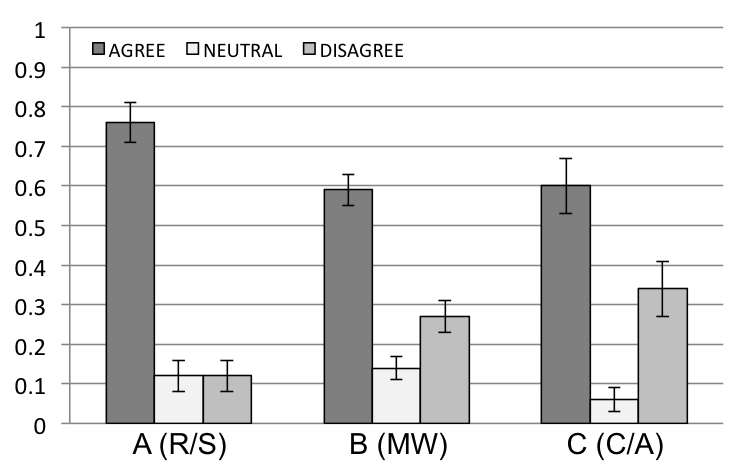}
\caption{Post-instruction student responses for courses A, B \& C.  Students indicated whether they agreed, disagreed, or felt neutral about the statement: \textit{When not being observed, an electron in an atom still exists at a definite (but unknown) position at each moment in time.} N = 69 (A), 135 (B) \& 47 (C); error bars represent the standard error on the proportion.}\label{fig:atom}
\end{figure}

These results stand in contrast to responses from the same students to the statement: \textit{When not being observed, an electron in an atom still exists at a definite (but unknown) position at each moment in time}.  A significant majority of the students from Instructor A's course expressed agreement with this statement; however, agreement was also the most common response in both of the other courses. [Fig. \ref{fig:atom}.]

\begin{figure}[b]
    \includegraphics[height=5.1cm, width=8.5cm]{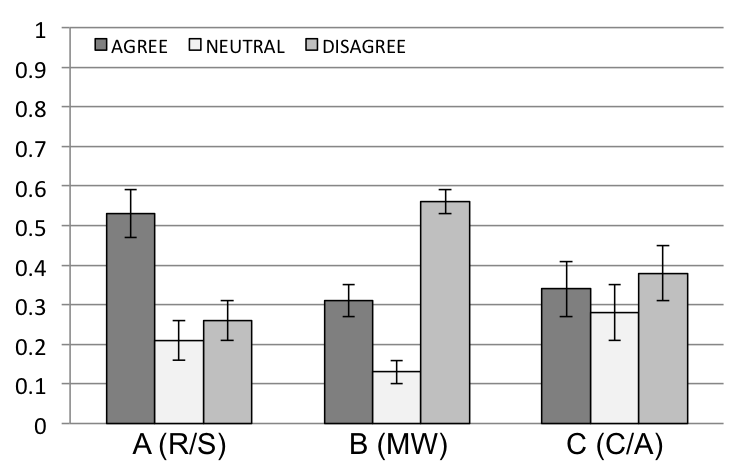}
\caption{Post-instruction student responses for courses A, B \& C.  Students indicated whether they agreed, disagreed, or felt neutral about the statement: \textit{The probabilistic nature of quantum mechanics is mostly due to the limitations of our measurement instruments.}  N = 69 (A), 135 (B) \& 47 (C); error bars represent the standard error on the proportion.}\label{fig:uncertain}
\end{figure}

Regardless of how one chooses to teach quantum physics, we believe most instructors would want their students to disagree with the statement: \textit{The probabilistic nature of quantum mechanics is mostly due to the limitations of our measurement instruments.}  For this statement,  students from Instructor A's course tended to agree, most of Instructor B's students preferred to disagree, and Instructor C's students were evenly split among the three possible responses.  [Fig. \ref{fig:uncertain}.]

\begin{figure}[t]
    \includegraphics[height=5.1cm, width=8.5cm]{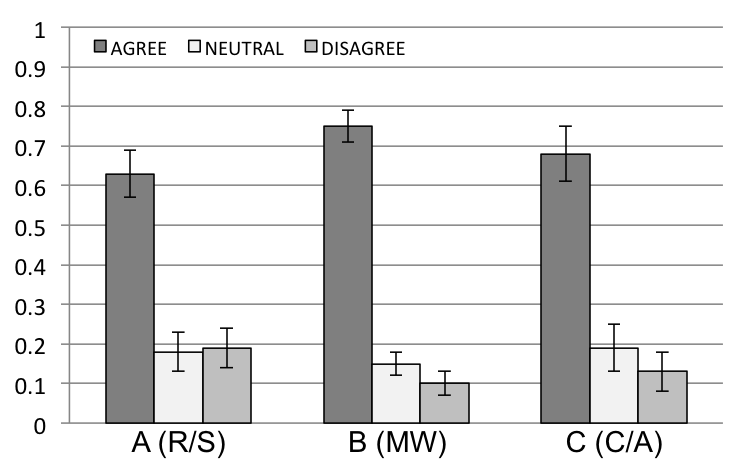}
\caption{Post-instruction student responses for courses A, B \& C.  Students indicated whether they agreed, disagreed, or felt neutral about the statement: \textit{I think quantum mechanics is an interesting subject.}  N = 69 (A), 135 (B) \& 47 (C); error bars represent the standard error on the proportion.}\label{fig:interest}
\end{figure}

In addition to learning course content, the promotion of student interest in quantum physics is also a common goal of instruction.  We measured this via responses to the statement: \textit{I think quantum mechanics is an interesting subject.} [Fig. \ref{fig:interest}.]  There is some variance between the three courses at post-instruction, but these differences are not statistically significant ($\chi^2 (4)=3.05, p=0.55$).

\subsection{\label{discussdata}Discussion}

The results presented above demonstrate that different instructional approaches with respect to interpretation can have different, measurable impacts on student thinking.  Moreover, they illustrate the contextual nature of students' conceptions of quanta, and imply that within specific contexts those conceptions are influenced most by explicit instruction.

Instructors A and B both taught their own physical interpretations of the double-slit experiment, and the most common responses from their respective students are aligned with that instruction.  At the same time, there was no bias among Instructor C's students towards any particular stance, which would be consistent with his approach if one were to characterize it as \textit{not} teaching any particular interpretation.  This result by itself is not sufficient to establish a direct link between this survey outcome and an instructor's lack of emphasis on interpretation, but similar results have been seen in the past in other C/A courses taught at CU \cite{baily2010teaching}.

Only Instructor A discussed his interpretation of atomic electron orbitals during lecture, and the post-instruction responses from his students are consistent with that instruction.  Neither Instructors B nor C brought up interpretive issues when teaching the Schr{\"o}dinger model of hydrogen, and the post-instruction responses from their students demonstrate a similar, though less strong, bias towards thinking of them as localized particles.

Our conclusions about the contextual nature of student thinking are further supported by our validation interviews, which indicated that students frequently modify their conceptions of quanta in a piecewise manner, both within and across contexts, often without looking for or requiring internal consistency.  Even when their instructors de-emphasized interpretation (explicitly or otherwise), students still developed a variety of ideas about quantum phenomena, some of which were highly nuanced, and others that emerged spontaneously as a form of sense making \cite{baily2010refined}.

The results for the statement about the probabilistic nature of quantum mechanics are reminiscent of those for the double-slit experiment essay question, in that the outcomes for courses A \& B were consistent with the interpretive approaches of their respective instructors.  The majority of students from the R/S course agreed with a statement that implies the use of probabilities to describe measurement outcomes stems from classical ignorance, whereas students from the MW course were most likely to disagree.  Instructor C's students were again, within statistical error, evenly split among the three possible responses.

As for student interest, we note that in each case at least a quarter of students chose \textit{not} to agree that quantum mechanics was interesting to them after a semester of instruction.  For all three courses, the most common reasons provided for giving a negative response were not perceiving the relevance of quantum physics to the macroscopic world, or to their training as engineers.  Among all the students' responses for each course, very few (if any) specifically mentioned the teaching style or the structure of the course as having influenced their opinion, whether positive or negative; however, this does not necessarily mean these factors had no impact on student affect.

Although we have not presented pre-instruction data in this section, these cohorts represent similar student populations, and the available data indicate there are no statistically significant differences between them at the beginning of the semester in terms of aggregate responses to these same survey statements.  As demonstrated in the next section, these three courses are not similar in terms of the ways in which students shifted in their responses between pre- and post-instruction.  We compare these shifts with those from two additional courses that used a curriculum designed to help students transition away from local realist interpretations of quantum phenomena, as well as promote greater interest in quantum physics.

\section{\label{sec:curriculum}Curriculum Development and Outcomes}

\begin{table*}[t]
\begin{tabular}{ p{5cm} p{10cm} }
   \hline
   Section & Topics \\
   \hline
   \multirow{4} {5cm} {I. Classical and Semi-classical Physics (Lectures 1-14)}
    & Introduction, review of mathematics and classical E\&M \\
    & Properties of waves, Young's double-slit experiment \\
    & Photoelectric effect, photons, polarization \\
    & Atomic spectra, lasers, Bohr model \\
   \hline
   \multirow{4} {5cm} {II. Development of Quantum Theory (Lectures 15-25)}
    & Atomic spin, Stern-Gerlach experiments, probabilistic measurements \\
    & EPR, entanglement, Local Realism, Complementarity \\
    & Single-photon experiments, electron diffraction, wave-particle duality \\
    & Wave functions, uncertainty principle, Schr{\"o}dinger equation \\    
   \hline
   \multirow{5} {5cm} {III. Applications of Quantum Mechanics (Lectures 26-40)}
    & Infinite and finite square wells \\
    & Tunneling, STM's, alpha decay\\
    & Hydrogen atom, periodic table, molecular bonding \\
    & Conductivity, semiconductors, diodes, transistors \\
    & Spin statistics, BEC, MRI \\
   \hline
\end{tabular}
\caption{Topics covered in a modern physics course that emphasized the physical interpretation of quantum mechanics.}\label{tab:outline}
\end{table*}

Informed by our research, we developed a new curriculum that had multiple aims, among them: (i) make the physical interpretation of quantum physics a topic unto itself, and consistently attend to student interpretations throughout the course; (ii) help students acquire the language and resources to identify and articulate their own (often unconscious) beliefs about reality and the nature of science; and (iii) provide experimental evidence that directly confronts their intuitive expectations.  Although we decided to promote a Matter-Wave perspective in this class, students were in no way evaluated based on their preferred interpretations.  During in-class discussions, we did not tell students they were necessarily wrong to make use of their classical intuitions as a form of sense making, though we did our best to demonstrate that local realist theories cannot reproduce all the predictions of quantum mechanics.  Our ultimate goal was for students to be able to perceive the distinctions between different perspectives, to recognize the advantages and limitations of each, and to apply this knowledge in novel situations.

\subsection{\label{course}Course Overview}

As with the other modern physics courses for engineering majors described above, ours spanned a 15-week semester, and consisted of large lectures meeting three times per week.  There were twice-weekly problem-solving sessions staffed by the authors (acting as co-instructors) and two undergraduate Learning Assistants \cite{otero2010la}, who also helped facilitate student discussion during lectures.  A total of 13 weekly homework assignments consisted of online submissions and written, long-answer questions; there was a broad mixture of conceptual and calculation problems, both requiring short-essay, multiple-choice, and numerical answers.  We gave three midterm exams outside of class, and there was a cumulative final.  At the end of the semester, in lieu of a long-answer section on the final exam, students wrote a 2-3 page (minimum) essay on a topic of their choice, or a personal reflection on their experience of learning about quantum mechanics in our class (an option chosen by $\sim$40\% of students).

Following the lead of the original course transformations, we omitted special relativity to win time for new material, which was mostly placed in the middle of the course.  The progression of topics can be broken into three main parts: (I) classical and semi-classical physics; (II) the development of quantum theory; and (III) its application to physical systems.  A detailed explication of this new curriculum and associated course materials \cite{baily2011thesis,modernmat} is beyond the scope of this article, but a summary overview of the topic coverage can be found in Table \ref{tab:outline}.

We augmented a number of standard topics (e.g., the uncertainty principle, atomic models) with interpretive discussions that had been missing in prior courses, and introduced several new topics (e.g., entanglement, single-photon experiments) that created additional opportunities for students to explore the differences between theory, experimental data, and the physical interpretation of both.  We took a `spins first' approach to Section (II) of this curriculum by starting with two-level systems before moving on to wave mechanics.  We consider the mathematical tools used in the former to be less complicated than those of the latter, such that concepts can be explored without the need for lengthy calculations.

The new material in Section (II) was drawn from a variety of sources, such as monographs \cite{baggott1992meaning,bell2004speak,greenstein2006challenge}, textbooks \cite{styer2000strange,scarani2010pieces}, journal articles \cite{tonomura1989electron,frabboni2012buildup} and popular science writing \cite{shimony1988reality,mermin1985moon}.  There were no textbooks covering all of the relevant material, so we used a combination of Vols. 3 \& 5 of Knight \cite{knight2004physics}, supplemented by other level-appropriate readings.  An online discussion board was created so that students could anonymously post questions about these readings and provide answers to each other, which granted us ample opportunity to gauge how students were responding to topics that are not a part of the standard curriculum.

\begin{figure}
\begin{framed}
    \includegraphics[height=1.5in, width=3in]{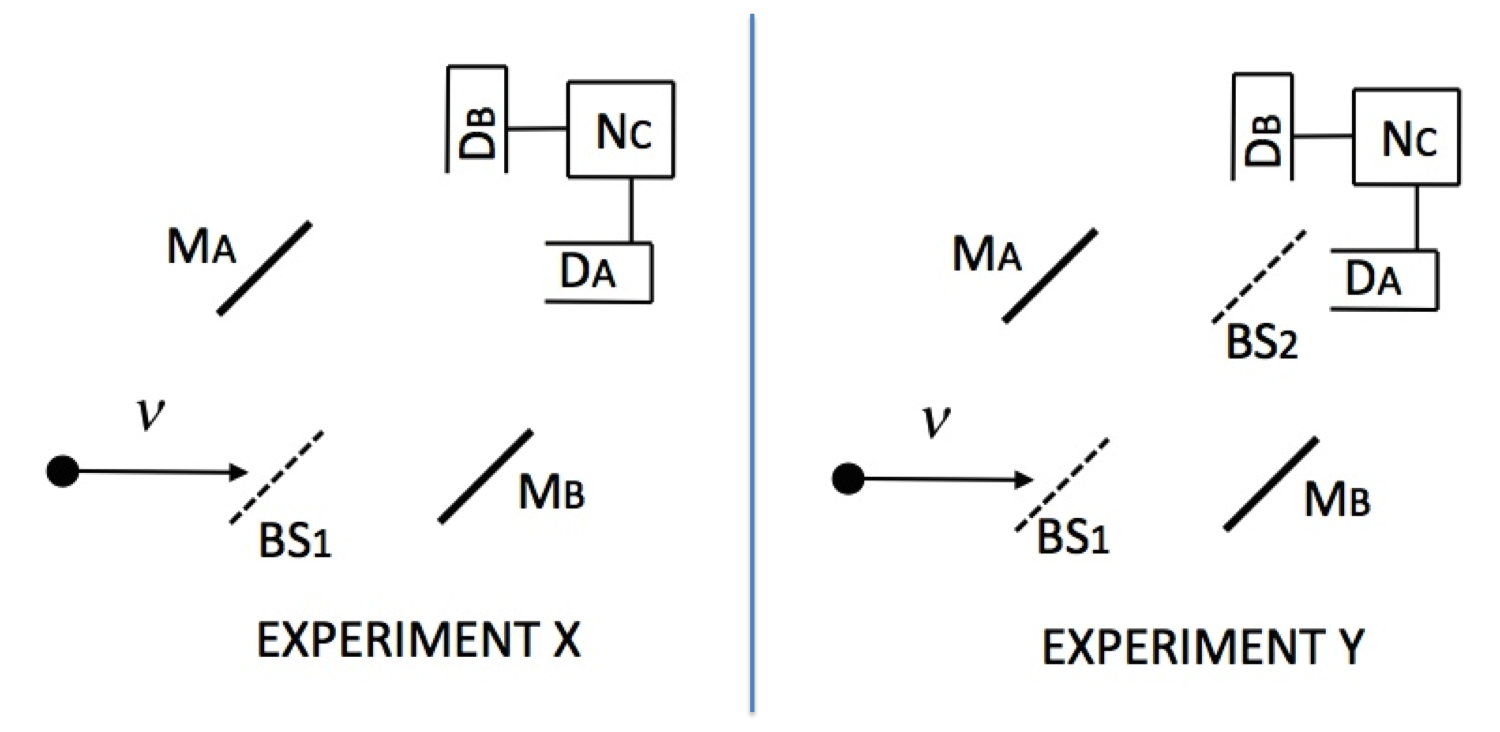}
\end{framed}
\caption{In each of these two experiments X (one beam splitter) \& Y (two beamsplitters), a single photon ($\nu$) is sent to the right through the apparatus. M = Mirror, BS = Beam Splitter, D = Detector, N$_{C}$ = Coincidence Counter.}\label{fig:experimentxy}
\end{figure}

One of our guiding principles was to present (as much as possible) experimental evidence that either supported or refuted different interpretations of quantum theory.  To illustrate how the topic of single-photon experiments \cite{grangier1986photon,greenstein2006challenge} contributed to this objective, consider Fig. \ref{fig:experimentxy}, which depicts an idealized single-photon experiment involving a Mach-Zehnder interferometer.  When just a single beam splitter is present (Experiment X), each photon is recorded in either one detector or the other, but never both; this result is often interpreted as meaning each photon took just one of the two paths with 50/50 probability.  When a second beam splitter is present (Experiment Y), interference effects can be observed by modulating the path length in just one of the arms of the interferometer.  This result can be interpreted as meaning each photon took both paths simultaneously, even though they are individually recorded in just one of the two detectors, for how can a change in just one of the paths otherwise effect the behavior of a photon that had supposedly only taken the other?

Some physicists would say that whether the second beam splitter is present or not determines whether the photon takes both paths or just one.  However, this explanation seems dubious in light of delayed-choice experiments \cite{jacques2007delay}, wherein the second beam splitter is either inserted or removed \textit{after} the photon has encountered the first beam splitter (the choice between configurations takes place outside the light cone of the photon's encounter with the first beam splitter).  Interference is observed if the second beam splitter is present, and otherwise not.

We taught our students that each photon always takes both paths simultaneously, regardless of whether the second beam splitter is present, as the most consistent way of interpreting the action of the beam splitter on the quantum state of the photon.  On the other hand, we felt that students should have multiple epistemological tools at their disposal, so we also explained that which type of behavior they should expect would depend on the ``path information'' available.  If it can be determined which path a photon had taken (from a realist perspective), there would be no interference; if not, then interference effects will be observed.  In doing so, we appealed to students' intuitions about classical particles (they are either reflected or transmitted) and classical waves (they are both reflected and transmitted).  Note that similar strategies can be employed with the double-slit experiment.

\begin{figure}[t]
   \includegraphics[height=4.7cm, width=7.05cm]{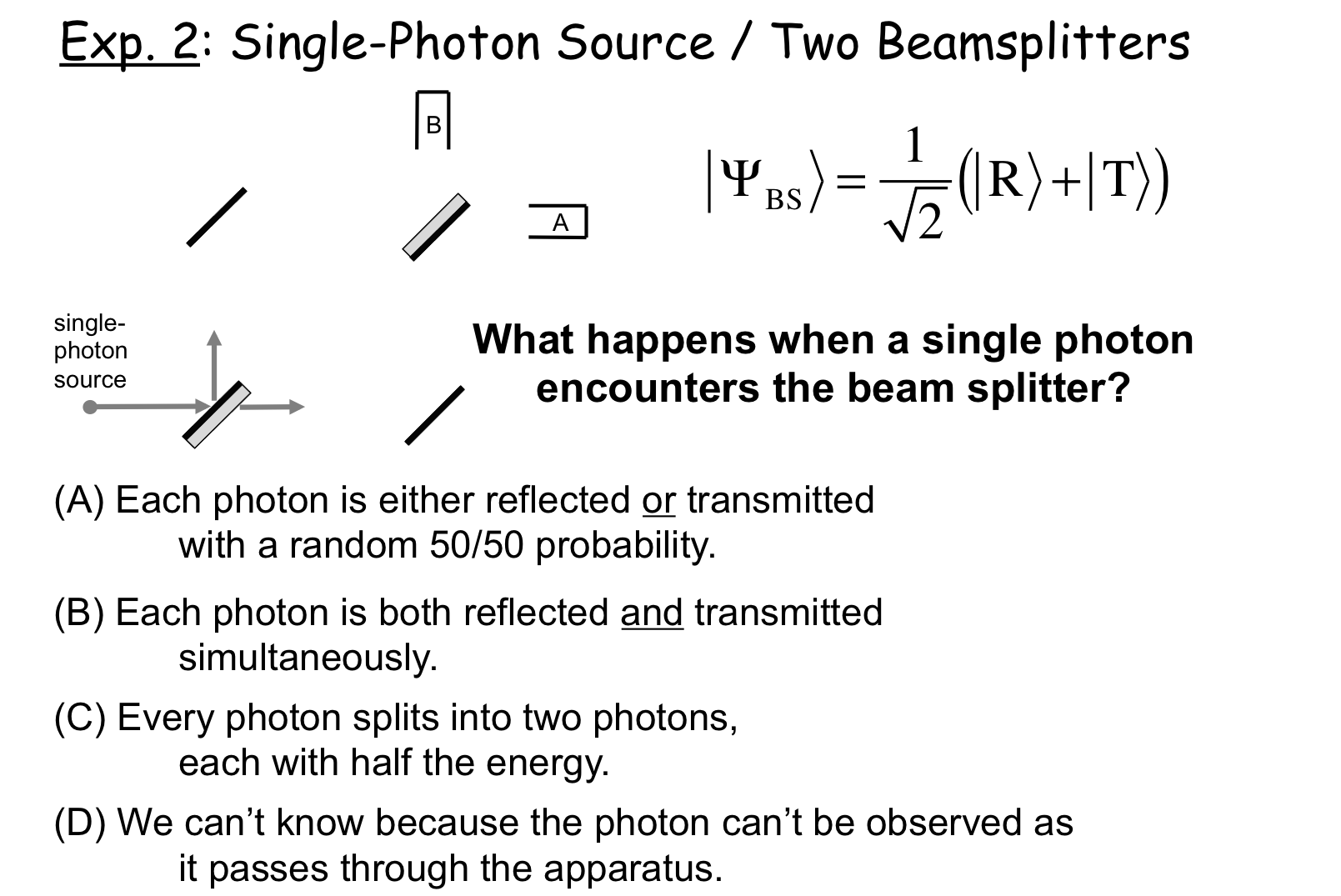}
\caption{Clicker question involving a Mach-Zehnder interferometer with a single-photon source, used during lecture to generate in-class discussion about physically interpreting experimental data and mathematical representations.}\label{fig:concept}
\end{figure}

These lectures were interspersed with clicker questions that prompted students to debate the implications of each experiment, and which provided an opportunity for them to distinguish between a collection of data points and an interpretation of what they signify.  It is important to emphasize that our interpretation-themed clicker questions generally did not have a single ``correct'' answer, such as the example shown in Fig. \ref{fig:concept} (which does contain at least one incorrect response).  The purpose of this question was to promote in-class discussion, and to elicit some of the ways students might interpret a mathematical representation of the photon's quantum state after encountering a beam splitter.  As instructors, we advocated for option (B) in this question, but we did not tell students who disagreed that their preferred perspective was necessarily incorrect.  As can be seen in this figure, one of the ways we made this topic more accessible to introductory students was to represent the state of the photon after the beam splitter as a superposition of the reflected and transmitted states, rather than the more technically correct description as entangled with the vacuum \cite{beck2012quantum}.

\subsection{\label{outcomes}Comparative Outcomes}

\begin{figure*}
   \includegraphics[height=10cm, width=14.5cm]{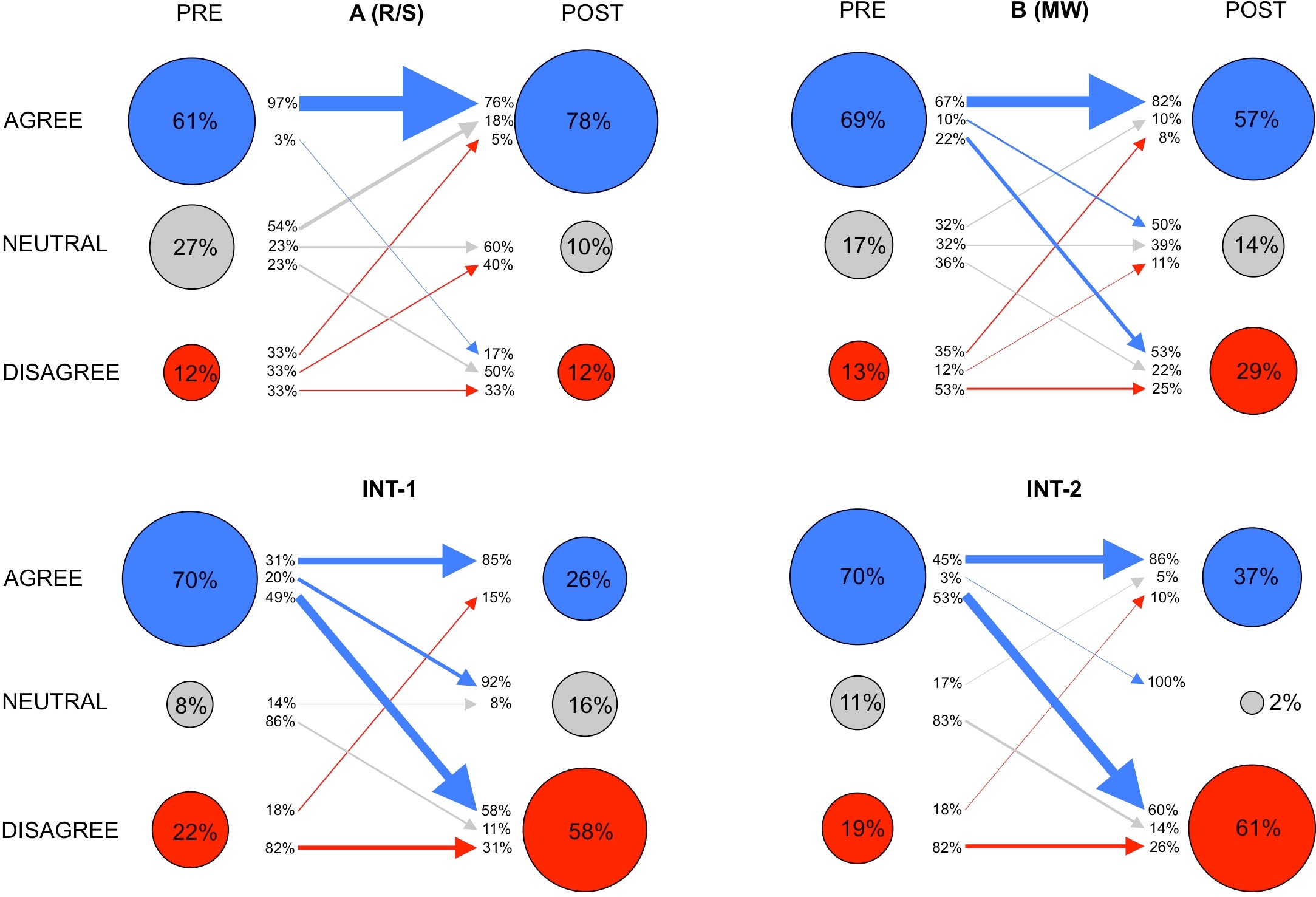}
\caption{Pre/post-instruction responses from four modern physics courses (as described in the text) to the statement: \textit{When not being observed, an electron in an atom still exists at a definite (but unknown) position at each moment in time.}  N = 49 (A), 126 (B), 77 (INT-1) \& 57 (INT-2).}\label{fig:Q3-All}
\end{figure*}

This new curriculum has thus far been implemented twice at CU (denoted here as INT-1 \& INT-2) with similar results, presented below in terms of pre- and post-instruction responses to the same three statements discussed in the previous section, from students in the R/S, MW, INT-1 and INT-2 courses.  Examining these shifts between the beginning and end of the semester further illustrates the differential impact of different instructional choices.  We were unable to collect pre-instruction data from Instructor C's course, but we can infer how his students' responses might have shifted if we assume their pre-instruction responses would have been similar to those from other modern physics courses for engineers.

In every case, results from the pre-instruction survey were not discussed with students, who were also not told they would be responding to the same survey questions at the end the course.  The pre/post-data sets below only represent students for whom we were able to match pre- and post-instruction responses, and not the full set of responses.  Table \ref{tab:response} shows the total number of students enrolled in each course at the beginning of the semester, the number of pre- and post-instruction survey response, and the number of matched pre/post responses.  For every course, and for each statement, the distributions for matched responses are statistically indistinguishable from the full pre- and post-data sets.

\begin{table}[t]\centering
\begin{tabular}{ | x{1.5cm} | x{1.5cm} | x{1.5cm} | x{1.5cm} | x{1.5cm} | }
   \hline
   \textbf{Course} & \textbf{Enrol.} & \textbf{Pre} & \textbf{Post} & \textbf{Matched} \tabularnewline
   \hline
   A (R/S) & 94 & 59 & 69 & 49 \tabularnewline
   \hline
   B (MW) & 146 & 136 & 135 & 126 \tabularnewline
   \hline
   INT-1 & 106 & 93 & 91 & 77 \tabularnewline
   \hline
   INT-2 & 81 & 64 & 71 & 57 \tabularnewline
   \hline
\end{tabular}
\caption{Total number of students enrolled at the start of the semester for each course, along with the number of pre- and post-instruction responses to the online survey, and the number of pre/post-matched responses.}
\label{tab:response}
\end{table}

\begin{figure*}
    \includegraphics[height=10cm, width=14.5cm]{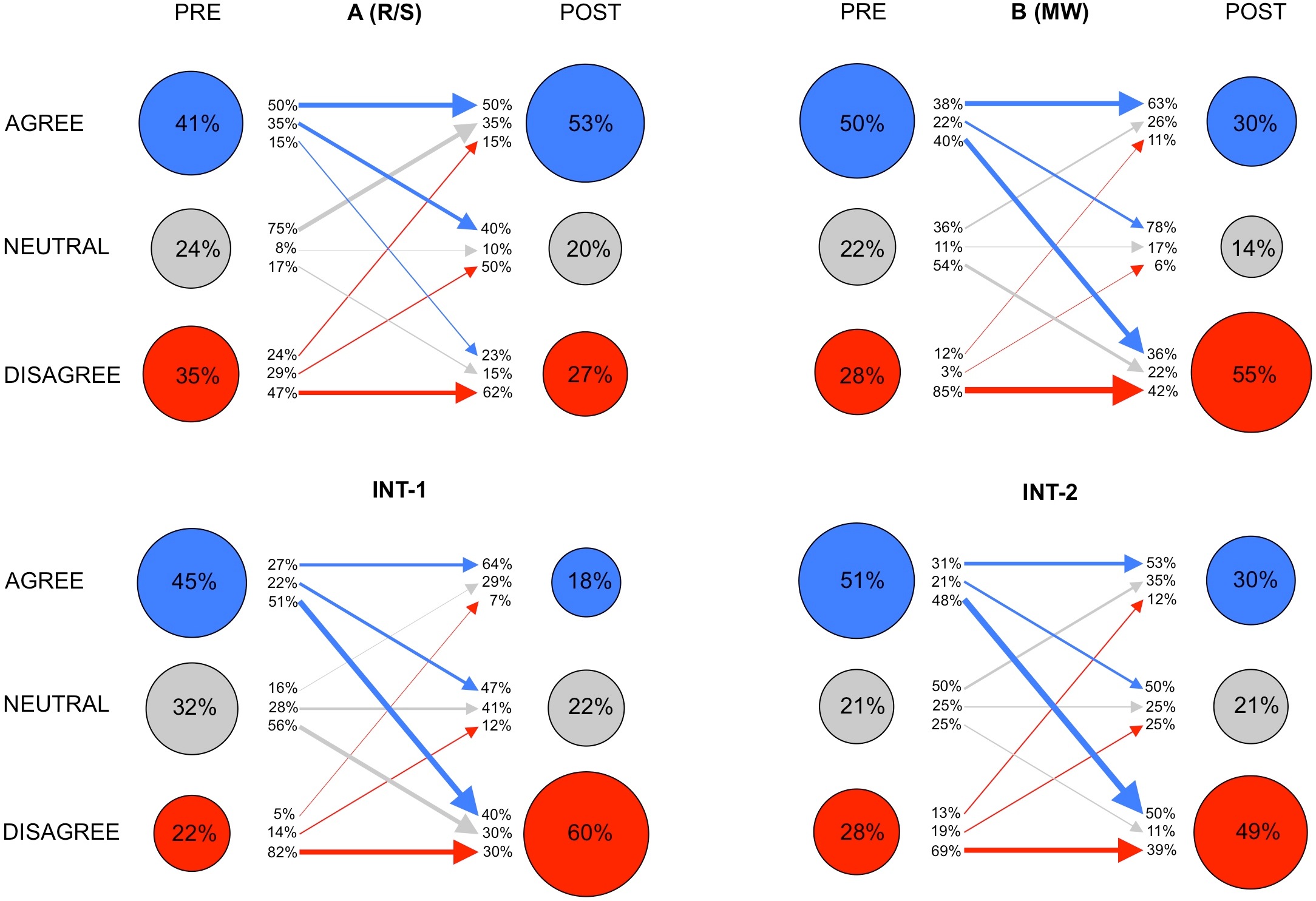}
\caption{Pre/post-instruction responses from four modern physics courses (as described in the text) to the statement: \textit{The probabilistic nature of quantum mechanics is mostly due to the limitations of our measurement instruments.}  N = 49 (A), 126 (B), 77 (INT-1) \& 57 (INT-2).}\label{fig:Q2-All}
\end{figure*}

\begin{figure*}
    \includegraphics[height=9.8cm, width=14.5cm]{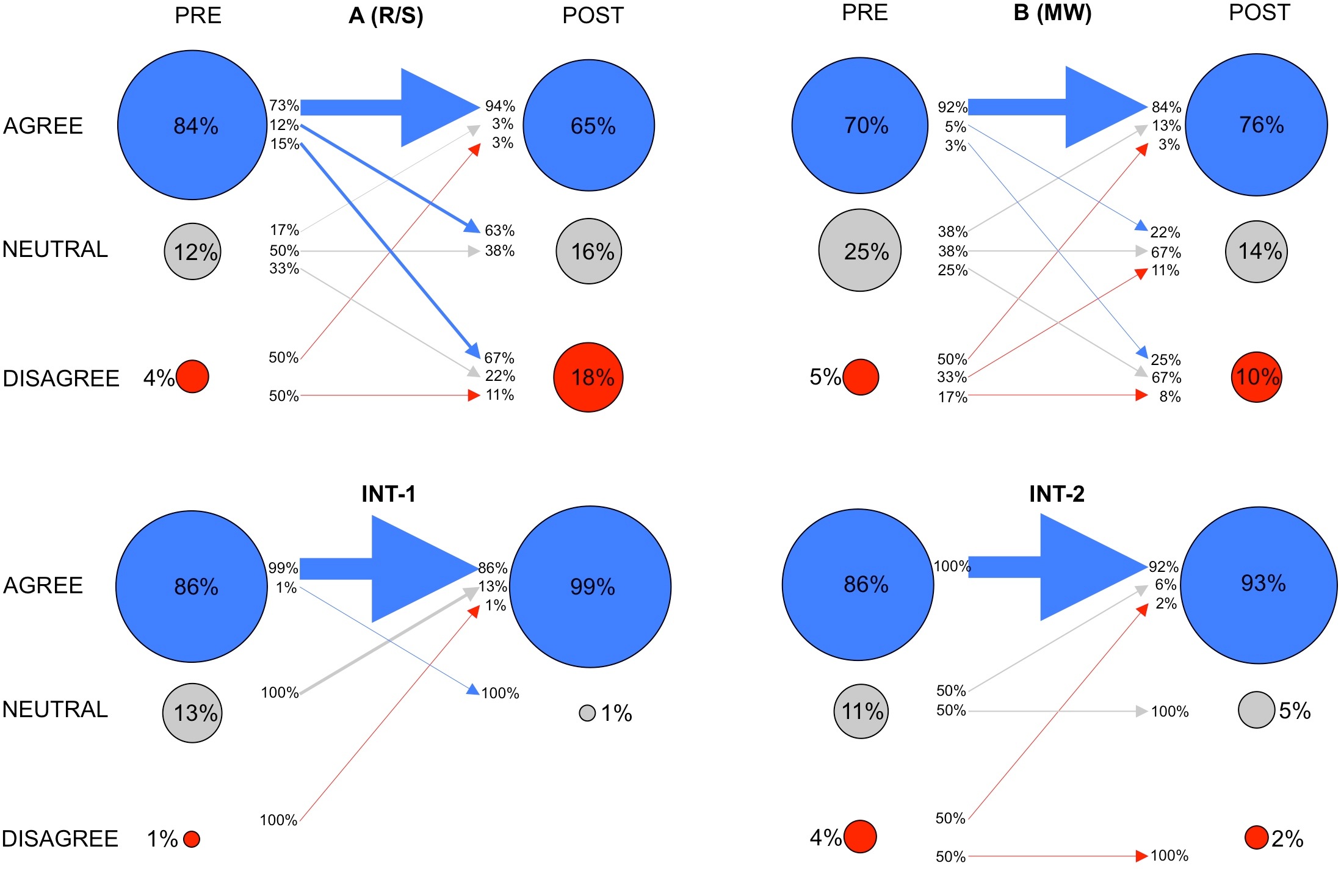}
\caption{Pre/post-instruction responses from four modern physics courses (as described in the text) to the statement: \textit{I think quantum mechanics is an interesting subject.}  N = 49 (A), 126 (B), 77 (INT-1) \& 57 (INT-2).}\label{fig:Q4-All}
\end{figure*}

In addition to aggregate pre/post comparisons, we also examine some of the dynamics in how students shifted between the beginning and end of the semester.  The visualizations shown in \Cref{fig:Q3-All,fig:Q2-All,fig:Q4-All} of these pre/post shifts (inspired by the discussion in Ref. \cite{wittmann2014plots}) reveal details that would have been lost if only the initial and final percentages were displayed.  For example, 12\% of students in the R/S course disagreed with the statement about atomic electrons at pre-instruction, and 12\% also at post-instruction, but these numbers do not represent the same groups of students.

For each of the four courses, the circles on the left side show the percentage of students who either agreed, disagreed or felt neutral about the given statement at the beginning of the semester, while the circles on the right show the same at post-instruction.  The area of each circle is proportional to the percentage of the total matched responses for that course.  In the space between these two sets of circles, the three numbers associated with each circle on the left represent the percentage of pre-instruction students in that group who shifted to each of the three post-instruction responses, and the thickness of each arrow is proportional to the percentage of students involved in that shift (relative to the total number of matched responses for that course).  The three numbers associated with each circle on the right represent the percentage of students in that post-instruction group who came over from each of the pre-instruction groups.

As a concrete example, for the R/S course shown in Fig. \ref{fig:Q3-All} (Course A, upper-left corner), at the beginning of the semester 61\% of matched respondents agreed with the statement about atomic electrons, 27\% responded neutrally and 12\% disagreed.  Of the group that had disagreed with the statement at pre-instruction, a third of them still disagreed at post-instruction, a third switched from disagreement to agreement, and the remaining third responded neutrally at the end of the semester.  Of the students who disagreed at post-instruction (also 12\% of the matched responses), 33\% had disagreed at the beginning of the semester, 50\% had originally responded neutrally and 17\% had switched from agreement to disagreement.

We first note that for all four courses the pre-instruction responses to the atomic electrons statement are roughly equivalent; the differences between the four are not statistically significant at the $p < 0.05$ level by a $\chi^2$ test ($p = 0.07$).  Almost every student in Course A who had agreed at the start of the semester still agreed at the end, the majority of those who had been neutral switched to agreement, as well as a third of those who had initially disagreed; there were fluctuations between responses, but the movement was predominantly towards the upper right (agreement).  For Course B, two thirds of the students who had agreed at pre-instruction also agreed at post-instruction, though a greater percentage of that group shifted towards disagreement than for Course A.  For the INT-1 \& 2 courses, the dominant tendency is a shift toward the lower right (disagreement).  Note also that, although the percentage of neutral responses for INT-1 increased over the semester, most of those neutral post-instruction responses were from students who had initially agreed with the statement, and most of those who had at first responded neutrally switched over to disagreement.

Fig. \ref{fig:Q2-All} shows pre/post responses and shifts for the statement about the probabilistic nature of quantum mechanics; again, the pre-instruction differences are not statistically significant ($p = 0.54$).  The post-instruction distributions for all four courses are significantly different ($p = 0.001$), but the differences between courses B, INT-1 \& 2 are not ($p = 0.24$).  As with the atomic electrons statement, the greatest tendency for Course A was a shift towards the upper right (agreement); also, most of those who felt neutral at the end of the semester had switched from other categories, and most who were initially neutral changed to agreement.  On the other hand, a shift towards post-instruction disagreement is predominant for the other three.

Pre-instruction responses for the four courses regarding student interest are not significantly different ($p = 0.06$), but the post-instruction distributions are ($p < 0.00001$). [Fig. \ref{fig:Q4-All}.]  Student interest in quantum mechanics decreased for Course A, and though the percentage expressing interest did increase in Course B, both A and B are similar in terms of the amount of ``cross-hatching" visible in the respective diagrams.  Remarkably, virtually every INT-1 student agreed at the end of the semester that quantum mechanics is interesting, and only one student switched from agreement to neutral.  For the INT-2 course, not a single student reported a decrease in their interest in quantum mechanics, and every student who initially agreed continued to do so at the end of the semester.

Although the post-instruction interest in quantum mechanics for INT-1 \& 2 is significantly greater than for Course B, the differential impact of these two types of instruction is less obvious because pre-instruction interest was lower for Course B, and student interest did increase in that course.  The difference is more apparent if we unpack the agreement category into \textit{agreement} and \textit{strong agreement}.  Table \ref{tab:strong} shows for each course the percentage of all matched students who either agreed or strongly agreed at pre- and post-instruction.  For the MW course, those numbers remained essentially the same, whereas students in the INT-1 \& 2 courses became more emphatic in their agreement that quantum mechanics is an interesting subject.  We conclude that this new curriculum was not only successful in maintaining student interest, but in promoting it as well.

\begin{table}[b]\centering
\begin{tabular}{ | x{1.5cm} | x{1.5cm} | x{1.5cm} | x{1.5cm} | x{1.5cm} | }
   \hline
    \textbf{ } & \multicolumn{2}{| c |}{\textbf{Pre (\%)}} & \multicolumn{2}{| c |}{\textbf{Post (\%)}}  \tabularnewline
   \cline{2-5}
   \textbf{Course} & Agree & Strongly Agree & Agree & Strongly Agree \tabularnewline
   \hline
   A (R/S) & 35 & 49 & 25 & 40 \tabularnewline
   \hline
   B (MW) & 31 & 39 & 35 & 41 \tabularnewline
   \hline
   INT-1 & 32 & 53 & 20 & 78 \tabularnewline
   \hline
   INT-2 & 16 & 70 & 7 & 86 \tabularnewline
   \hline
\end{tabular}
\caption{Percentage of matched students from each course who at pre- and/or post-instruction either agreed or strongly agreed with the statement: \textit{I think quantum mechanics is an interesting subject}}\label{tab:strong}
\end{table}

\section{\label{sec:conclusion}Summary and Discussion}

We have frequently heard that a primary goal when introducing students to quantum mechanics is for them to recognize a fundamental difference between classical and quantum uncertainty.  The notorious difficulty of accomplishing this has led many instructors to view this learning goal as superficially possible, but largely unachievable in a meaningful way for most undergraduate students \cite{dubson2009faculty}.  We believe our studies demonstrate otherwise.  By making questions of classical and quantum reality a central theme of our course, and also by making their own beliefs (and not just those of scientists) a topic of discussion, we were able to positively influence student thinking across a variety of measures.  We have presented data from several particular courses, but the results reported here for the R/S, MW \& C/A courses are typical of other, similar courses that have been discussed elsewhere \cite{baily2010teaching,baily2010hidden}.

The outcomes for Instructor A's course were generally aligned with his instructional approach: electrons are localized entities, and quantum uncertainty is not much different from classical ignorance.  While this is not a particularly common way of teaching quantum physics, there have been other instances at CU of a similar approach being taken, and we suspect this also occurs at other institutions, and at a variety of levels of instruction.  Understanding how this approach can impact student thinking is therefore important, particularly when it may negatively impact student affect.

We characterized Instructor B's course as having explicitly taught an MW interpretation of the double-slit experiment (though not framed as an interpretation), but then de-emphasized interpretive themes in the latter stages of the semester.  This is also reflected in the outcomes for his course, in that students were likely to have adopted his perspective in a context where the instruction had been explicit, but much less likely in another context where it was not.  The MW approach did result in significant shifts in student perspectives on the nature of quantum uncertainty (on par with the INT-1 \& 2 courses), but was less successful than ours in promoting and maintaining student interest.

With regard to the double-slit essay question and the statement about the probabilistic nature of quantum mechanics, Instructor C's approach resulted in the greatest mixture of post-instruction responses, evenly distributed across the three perspectives.  The post-instruction distribution for the statement about atomic electrons is essentially identical to the results from the MW course.  If we assume the pre-instruction responses would have been similar to those for other engineering courses, the C/A approach had little impact on students' ideas about atomic electrons, was not as successful as the MW \& INT courses at influencing their perspectives on quantum uncertainty, and resulted in decreased interest in quantum mechanics.

Even though Instructor B's approach to interpretation differed in obvious ways from Instructor C's, it turns out that pragmatism was also a motivating factor in his instructional choices.  Because de-emphasizing the physical interpretation of quantum mechanics is so common, it is worthwhile to consider some of the reasons for this in greater detail, as explained by Instructor B in an interview at the end of the semester:

\begin{quote}``This [probabilistic] aspect of quantum mechanics I feel is very important, but I don't expect undergraduate students to grasp it after two months.  So that's why I can understand why [the survey statement about atomic electrons] was not answered to my satisfaction, but that was not my primary goal of this course, not at this level.  We don't spend much time on this introduction to quantum mechanics, and there are many aspects of it that are significant enough at this level.  It is really great for students to understand how solids work, how does conductivity work, how does a semiconductor work, and these things you can understand after this class.  If all of the students would understand how a semiconductor works, that would be a great outcome.  I feel that probably at this level, especially with many non-physics majors, I think that's more important at this point.

But still, they have to understand the probabilistic nature of quantum mechanics, and I hope, for instance, that this is done with the hydrogen atom orbitals {-} not that everyone would understand that, but if the majority gets it that would be nice.  These are very hard concepts.  At this level, I feel it should still have enough connections to what they already understand, and what they want to know.  They want to know how a semiconductor works, probably much more than where is an electron in a hydrogen atom.

I don't think the [engineering] students will be more successful in their scientific endeavors, whether it's a personal interest or career, by giving them lots and lots of information about how to think of the wave function.  The really important concept I feel is to see that there is some sort of uncertainty involved, which is new, which is different from classical mechanics. [...] At the undergraduate level, I feel it is important to make the students curious to learn more about it, and so even if they don't understand everything from this course, if they are curious about it, that's more important than to know where the electron really is, I think.'' \end{quote}

To summarize, Instructor B felt that understanding the nature of uncertainty in quantum mechanics \textit{is} an important learning goal, but one that will likely not be achieved by many students at this level.  He \textit{assumed} engineering students would be more interested in the practical aspects of quantum physics.  He said he would have \textit{liked} for his students to disagree with the idea of localized atomic electrons, and yet $\sim$75\% of them chose to \textit{not} disagree at the end of the semester.

If the aim of instruction is not necessarily a complete understanding of the concepts, but for students to at least come away with a continued interest in quantum physics, then we would claim the INT-1 \& 2 courses were more successful in this regard.  We should also not presume to know exactly where the interests of our students lie.  The results from our implementations suggest that engineering students were in fact \textit{just as interested} (if not more so) in contemplating the nature of reality and learning about applications of entanglement to quantum cryptography as they were in learning about semiconductors.  And finally, our students \textit{did} learn about semiconductors, as well as conduction banding, transistors and diodes.

Although transitioning students away from classical perspectives was one of our goals, we would not connote too much negativity with students relying on their intuition as a form of sense making.  Indeed, our approach to teaching quantum interpretations frequently involved an appeal to students' understanding of classical systems (e.g., particles are either transmitted or reflected; they are localized upon detection), which in fact is consistent with the Copenhagen interpretation.  Everyday thinking can be misleading in quantum physics, but that is not a sufficient argument for the wholesale abandonment of productive epistemological tools.  What is important is that students understand the limitations of these intuitive conceptions, and where they might lead them astray.

Just as important is the recognition that most modern physics curricula ignore the fact that a ``second quantum revolution'' has taken place in the last decades, due to the realization of single-quanta experiments, and a corresponding appreciation of the significance of entanglement \cite{aspect2004bell}.  Ideas that were once relegated to the realm of metaphysics are now driving exciting areas of contemporary research, and it is possible to make these developments accessible to introductory quantum physics students.

\begin{acknowledgments}

This work was supported in part by NSF CAREER Grant \# 0448176, NSF DUE \# 1322734, the University of Colorado and the University of St Andrews.  Particular thanks go to the students and instructors who made these studies possible.

\end{acknowledgments}

\bibliography{references}
\bibliographystyle{apsper}   

\end{document}